\documentclass[aps,twocolumn,superscriptaddress,showpacs]{revtex4}

\usepackage[dvips]{graphicx}
\usepackage{color}  
\newcommand{\nc}{\newcommand}           
\nc{\vc}[1]     {\mbox{\boldmath $#1$}} 
\nc{\mapleft}[1]{                       
 \smash{\mathop{                      %
  \hbox to 0.90cm{\rightarrowfill} }\limits_{#1}}}

\nc{\mydraft}	{\setlength{\topmargin}{-1.5cm}}
\mydraft

\begin{document}

\title{Five-body resonances of $^8$C using the complex scaling method}

\author{Takayuki Myo\footnote{myo@ge.oit.ac.jp}}
\affiliation{General Education, Faculty of Engineering, Osaka Institute of Technology, Osaka, Osaka 535-8585, Japan}
\affiliation{Research Center for Nuclear Physics (RCNP), Osaka University, Ibaraki 567-0047, Japan}

\author{Yuma Kikuchi\footnote{yuma@rcnp.osaka-u.ac.jp}}
\affiliation{Research Center for Nuclear Physics (RCNP), Osaka University, Ibaraki 567-0047, Japan}

\author{Kiyoshi Kat\=o\footnote{kato@nucl.sci.hokudai.ac.jp}}
\affiliation{Division of Physics, Graduate School of Science, Hokkaido University, Sapporo 060-0810, Japan}

\date{\today}

\begin{abstract}
We study the resonance spectroscopy of the proton-rich nucleus $^8$C in the $\alpha$+$p$+$p$+$p$+$p$ cluster model.
Many-body resonances are treated on the correct boundary condition as the Gamow states using the complex scaling method.
We obtain the ground state of $^8$C as a five-body resonance for the first time, which has dominantly the sub-closed $(p_{3/2})^4$ configuration and agrees with the recent experiment for energy and decay width. 
We predict the second $0^+$ state with the excitation energy of 5.6 MeV, which corresponds to the $2p2h$ state from the ground state. 
We evaluate the occupation numbers of four valence-protons in the $^8$C states and also the $J^\pi$ distribution of proton-pair numbers of the two $0^+$ states of $^8$C.
The ground state involves a large amount of the $2^+$ proton-pair fraction, while the excited $0^+_2$ state almost consists of two of the $0^+$ proton pairs,
which can be understood from the $(p_{3/2})^2(p_{1/2})^2$ configuration.
We also discuss the mirror symmetry between $^8$C and $^8$He with an $\alpha$+four nucleon picture.
It is found that the $0^+$ states retain the mirror symmetry well for the configuration properties of two nuclei.
\end{abstract}

\pacs{
21.60.Gx,~
21.10.Pc,~
21.10.Dr,~
27.20.+n~
}


\maketitle 

\section{Introduction}

Radioactive beam experiments have provided us with much information on unstable nuclei far from the stability.
In particular, the light nuclei near the drip-line exhibit new phenomena of nuclear structures,
such as the neutron halo structure found in $^6$He, $^{11}$Li and $^{11}$Be \cite{tanihata85}.
The unstable nuclei can often be unbound states beyond the particle thresholds due to the weak binding nature.
The resonance spectroscopy of unbound states beyond the drip-line has been developed experimentally.
In addition to the energies and decay widths, the configuration information is important to understand the structures of the resonances.
In proton-rich and neutron-rich nuclei, the configurations of extra nucleons provide with the useful information to know the correlations between
the extra nucleons in resonances as well as in weakly bound states.
It is interesting to compare the structures of resonances and weakly bound states between proton-rich and neutron-rich sides.
This comparison is related to the mirror symmetry in unstable nuclei having a large isospin.

Recently, the new experiments on $^8$C have been reported \cite{charity10,charity11} in addition to the old observations\cite{roberton74,roberton76,tribble76}.
The $^8$C nucleus is known as an unbound system beyond the proton drip-line and its ground state is naively considered to be the $0^+$ resonance.
The ground state of $^8$C is observed at 2 MeV above the $^6$Be+$2p$ threshold energy and is close to the
$^7$B+$p$ threshold\cite{charity11}, and excited states have not yet been observed.
The $^8$C states can decay not only to a two-body $^7$B+$p$ channel, but also to many-body channels of $^6$Be+$2p$, $^5$Li+3$p$ and $^4$He+4$p$. This multi-particle decay condition makes difficulty to identify the states of $^8$C experimentally.

The mirror nucleus of $^8$C is $^8$He with isospin $T=2$, which has the bound ground state.
Recently, many experiments on $^8$He have been reported \cite{iwata00,meister02,chulkov05,skaza07,golovkov09,tanihata92,mueller07}.
Its ground state is considered to have a neutron skin structure consisting of four valence neutrons around $^4$He with the small binding energy of 3.1 MeV.
For the excited states of $^8$He, most of them can be located above the $^4$He+4$n$ threshold energy \cite{skaza07}.
This fact indicates that the observed resonances of $^8$He can decay into 
the channels of $^7$He+$n$,  $^6$He+2$n$, $^5$He+3$n$ and $^4$He+4$n$.
These multi-particle decays of $^8$He are related to the Borromean nature of $^6$He, which breaks up easily into $^4$He+$2n$,
and make it difficult to settle the excited states of $^8$He.
There still remain contradictions in the observed energy levels of $^8$He.

From the view point of the ``$^4$He+four protons or four neutrons'' system,
the information of $^8$C and $^8$He is important to understand the structures on and outside the drip-lines as a five-body picture.
It is also interesting to examine the effect of Coulomb interaction and the mirror symmetry in two nuclei.
Structures of resonances and weakly bound states generally depend on the existence of the open channels as the thresholds of the particle emissions.
In this sense, the mirror symmetry in unstable nuclei can be related to the coupling behavior to the open channels.
In the previous analyses of $^7$B and $^7$He with the $^4$He+$N+N+N$ model \cite{myo117},
we discussed the mirror symmetry in two nuclei.
It is found that breaking of the mirror symmetry is occurred in their ground states with respect to the amount of the mixing of $2^+$ states of $A=6$ subsystems,
while their excited states retain the symmetry well. This result concerns with the relative energy positions between the $A$=$7$ states and the ``$A$=$6$''+$N$ thresholds.
Similarly, it is interesting for $^8$C and $^8$He to compare the effects of the couplings to the open channels in the resonances of two nuclei.
The configuration properties of the extra four nucleons in $^8$C and $^8$He are also interesting from the viewpoint of the correlations of extra nucleons.

On the theoretical side, to treat the unbound states explicitly, several methods have been developed,
such as the microscopic cluster model \cite{adahchour06,arai09}, the continuum shell model \cite{volya05} and the Gamow shell model \cite{betan09,michel07}.
It is, however, difficult to satisfy the multi-particle decay conditions correctly for all open channels. 
For $^8$C, it is necessary to describe the $^4$He+4$p$ five-body resonances in the theory.
So far, no theory describes the $^8$C nucleus as five-body resonances.
In addition, it is important to reproduce the threshold energies of subsystems for particle decays, namely, the positions of open channels.
Emphasizing these theoretical conditions, in this study, we employ the cluster orbital shell model (COSM) \cite{myo117,suzuki88,masui06,myo077,myo097,myo10} of the $^4$He+$4p$ five-body system.
In COSM, the effects of all open channels are taken into account explicitly\cite{myo077} so that we can treat the many-body decaying phenomena.
In our previous works of neutron-rich systems\cite{myo077,myo097,myo10}, we have successfully described 
the He isotopes with the $^4$He+$4n$ model up to the five-body resonances of $^8$He including the full couplings with $^{5,6,7}$He.
We have described many-body resonances using the complex scaling method (CSM) \cite{ho83,moiseyev98,aoyama06} under the correct boundary conditions for all decay channels. 
In CSM, the resonant wave functions are directly obtained by diagonalization of the complex-scaled Hamiltonian using the $L^2$ basis functions.
Results for light nuclei using CSM have been obtained successfully for energies, decay widths, spectroscopic factors and also for the breakup strengths induced by the Coulomb excitations\cite{myo01,myo0711}, 
monopole transition\cite{myo10} and one-neutron removal\cite{myo097}. 
Recently, CSM has been developed to apply to the nuclear reaction methods such as 
the scattering amplitude calculation \cite{kruppa07}, Lippmann-Schwinger equation\cite{kikuchi10,kikuchi11} and the continuum-discretized coupled-channel (CDCC) method\cite{matsumoto10}.

In this study, we proceed with our study of resonance spectroscopy of the proton-rich nucleus $^8$C with the $^4$He+$4p$ five-body cluster model.
This study is the extension of the previous one of $^7$B with the $^4$He+$3p$ model \cite{myo117}.
We concentrate on the $0^+$ states of $^8$C and discuss the structure differences between the ground and the excited states.
It is interesting to examine how our model describes $^8$C as five-body resonances. 
We predict the resonances of $^8$C and investigate their binding properties.
To extract the information of the extra protons, we calculate the $J^\pi$ distribution of the pair numbers of the four valence protons in the $^8$C states.
This quantity is useful for understanding the coupling behavior of four protons as a proton-pair inside $^8$C.
For mirror nucleus $^8$He, we have performed the similar analysis\cite{myo10}, in which the large mixing of the $2^+$ neutron pair is confirmed for the ground state.
From the viewpoint of the mirror symmetry, we compare the structures of $^8$C with those of $^8$He and discuss the similarity and the difference in two nuclei.

In Sec.~\ref{sec:model}, we explain the complex-scaled COSM wave function.
In Sec.~\ref{sec:result}, we discuss the structures and the configurations of four valence protons in the ground and the excited states of $^8$C.
A Summary is given in Sec.~\ref{sec:summary}.

\section{Complex-scaled COSM}\label{sec:model}

\subsection{COSM for the $^4$He+\vc{N_{\rm v} p} systems}

We use COSM of the $^4$He+$N_{\rm v} p$ systems, where $N_{\rm v}$ is a valence proton number around $^4$He, 
namely, $N_{\rm v}=4$ for $^8$C.
The Hamiltonian form is the same as that used in Refs.~\cite{myo077,myo097,myo117};
\begin{eqnarray}
	H
&=&	\sum_{i=1}^{N_{\rm v}+1}{t_i} - T_G + \sum_{i=1}^{N_{\rm v}} V^{\alpha p}_i + \sum_{i<j}^{N_{\rm v}} V^{pp}_{ij}
    \\
&=&	\sum_{i=1}^{N_{\rm v}} \left[ \frac{\vec{p}^2_i}{2\mu} + V^{\alpha p}_i \right] + \sum_{i<j}^{N_{\rm v}} \left[ \frac{\vec{p}_i\cdot \vec{p}_j}{4m} + V^{pp}_{ij} \right] ,
    \label{eq:Ham}
\end{eqnarray}
where $t_i$ and $T_G$ are the kinetic energies of each particle ($p$ and $^4$He) and of the center of mass of the total system, respectively.
The operator $\vec{p}_i$ is the relative momentum between $p$ and $^4$He. 
The reduced mass $\mu$ is $4m/5$ using a nucleon mass $m$.
The $^4$He-$p$ interaction $V^{\alpha p}$ is given by the microscopic KKNN potential \cite{aoyama06,kanada79} for the nuclear part,
in which the tensor correlation of $^4$He is renormalized on the basis of the resonating group method in the $^4$He+$N$ scattering.
For the Coulomb part, we use the folded Coulomb potential using the density of $^4$He having the $(0s)^4$ configuration.
We use the Minnesota potential \cite{tang78} as the nuclear part of $V^{pp}$ in addition to the Coulomb interaction.
These interactions reproduce the low-energy scattering of the $^4$He-$N$ and the $N$-$N$ systems, respectively.

\begin{figure}[t]
\centering
\includegraphics[width=7.5cm,clip]{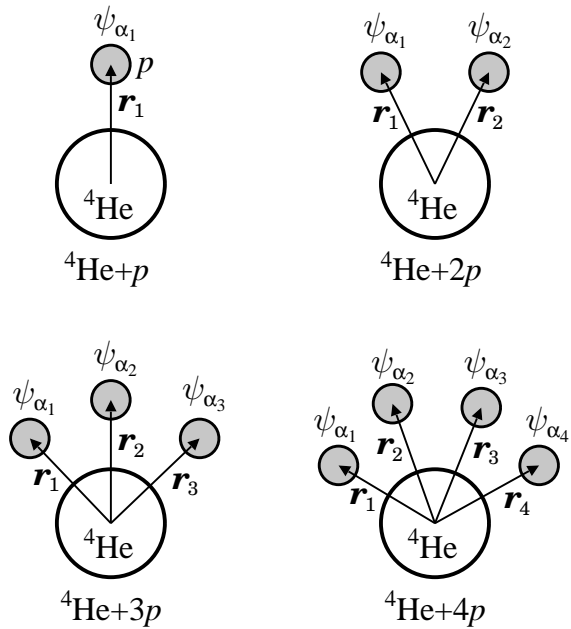}
\caption{Sets of the spatial coordinates in COSM for the $^4$He+$N_{\rm v} p$ system.}
\label{fig:COSM}
\end{figure}

For the wave function, $^4$He is treated as a $(0s)^4$ configuration of a harmonic oscillator wave function, 
whose length parameter is 1.4 fm to fit the charge radius of $^4$He as 1.68 fm.
The motion of valence protons around $^4$He is solved variationally using the few-body technique.
We expand the relative wave functions of the $^4$He+$N_{\rm v} p$ system using the COSM basis states \cite{suzuki88,masui06,myo077,myo097}.
In COSM, the total wave function $\Psi^J$ with spin $J$ is represented by the superposition of the configuration $\Psi^J_c$ as
\begin{eqnarray}
    \Psi^J
&=& \sum_c C^J_c \Psi^J_c,
    \label{WF0}
    \\
    \Psi^J_c
&=& \prod_{i=1}^{N_{\rm v}} a^\dagger_{\alpha_i}|0\rangle, 
    \label{WF1}
\end{eqnarray}
where the vacuum $|0\rangle$ is given by the $^4$He ground state.
The creation operator $a^\dagger_{\alpha}$ is for the single particle state of a valence proton above $^4$He
with the quantum number $\alpha=\{n,\ell,j\}$ in a $jj$ coupling scheme.
Here, the index $n$ represents the different radial component. 
The index $c$ represents the set of $\alpha_i$ as $c=\{\alpha_1,\cdots,\alpha_{N_{\rm v}}\}$.
We take a summation over the available configurations in Eq.~(\ref{WF0}), which gives a total spin $J$.
The expansion coefficients $\{C_c^J\}$ in Eq.~(\ref{WF0}) are determined variationally 
with respect to the total wave function $\Psi^J$ by the diagonalization of the Hamiltonian matrix elements.
The relation $\sum_{c} \left(C_c^J\right)^2=1$ is satisfied due to the normalization of the total wave function.

The coordinate representation of the single particle state corresponding to $a^\dagger_{\alpha}$ is given as 
$\psi_{\alpha}(\vc{r})$ as function of the relative coordinate $\vc{r}$ between the center of mass of $^4$He 
and a valence proton \cite{suzuki88}, as shown in Fig.~\ref{fig:COSM}.
We employ a sufficient number of radial bases of $\psi_\alpha(\vc{r})$ to describe the spatial extension of valence protons in the resonances, in which $\psi_\alpha(\vc{r})$ are normalized.
In this model, the radial part of $\psi_\alpha(\vc{r})$ is expanded with the Gaussian basis functions for each orbit as
\begin{eqnarray}
    \psi_\alpha(\vc{r})
&=& \sum_{k=1}^{N_{\ell j}} d^k_{\alpha}\ \phi_{\ell j}^k(\vc{r},b_{\ell j}^k),
    \label{WFR}
    \\
    \phi_{\ell j}^k(\vc{r},b_{\ell j}^k)
&=& {\cal N}\, r^{\ell} e^{-(r/b_{\ell j}^k)^2/2} [Y_{\ell}(\hat{\vc{r}}),\chi^\sigma_{1/2}]_{j},
    \label{Gauss}
	\\
    \langle \psi_\alpha|\psi_{\alpha'} \rangle 
&=& \delta_{\alpha,\alpha'}.
    \label{Gauss2}
\end{eqnarray}
The index $k$ is for the Gaussian basis with the length parameter $b_{\ell j}^k$.
Normalization factor of the basis and a basis number are given by ${\cal N}$ and $N_{\ell j}$, respectively. 
The coefficients $\{d^k_{\alpha}\}$ in Eq.~(\ref{WFR}) are determined using the Gram-Schmidt orthonormalization,
and hence the basis states $\psi_\alpha$ are orthogonal to each other as shown in Eq.~(\ref{Gauss2}).
The numbers of the radial bases of $\psi_\alpha$ are at most $N_{\ell j}$, and are determined to converge the physical solutions.
The same method using Gaussian bases as a single particle basis is employed in the tensor-optimized shell model\cite{myo09,myo11}.
The antisymmetrization between a valence proton and $^4$He is treated on the orthogonality condition model \cite{aoyama06}, 
in which the single particle state $\psi_{\alpha}$ is imposed to be orthogonal to the $0s$ state occupied by $^4$He.
The length parameters $b_{\ell j}^k$ are chosen in geometric progression \cite{myo097,aoyama06}.
We use at most 17 Gaussian basis functions by setting $b_{\ell j}^k$ from 0.2 fm to around 40 fm
with the geometric ratio of 1.4 as a typical one.
Due to the expansion of the radial wave function using a finite number of basis states, 
all the energy eigenvalues are discretized for bound, resonant and continuum states.
To obtain the Hamiltonian matrix elements of multi-proton system in the COSM configurations,
we employ the $j$-scheme technique of the shell model calculation in terms of $\psi_\alpha$ as the basis states.

In COSM, the asymptotic boundary condition of the wave functions for proton emissions is correctly described \cite{myo117,aoyama06,myo05}.
For $^8$C, all the channels of $^8$C, $^7$B+$p$, $^6$Be+$2p$, $^5$Li+$3p$ and $^4$He+$4p$ are automatically included in the total wave function $\Psi^J$ in Eq.~(\ref{WF0}).
These channels are coupled to each other by the interactions and the antisymmetrization,
and those couplings depend on the relative distances between $^4$He and a valence proton and between the valence protons.

We explain the parameters of the model space of COSM and the Hamiltonian which are determined in the previous analyses of He isotope\cite{myo077,myo097}.
For the single-particle states, we take the angular momenta $\ell\le 2$ to keep the accuracy of the converged energy within 0.3 MeV of $^6$He with the $^4$He+$n$+$n$ model in comparison with the full space calculation\cite{aoyama06}. 
In this model, we adjust the two-neutron separation energy of $^6$He($0^+$) to the experiment of 0.975 MeV 
by taking the 173.7 MeV of the repulsive strength of the Minnesota potential instead of the original value of 200 MeV.
The adjustment of the $NN$ interaction is originated from the pairing correlation between valence protons with higher angular momenta $\ell>2$ \cite{aoyama06}.
Hence, the present model reproduces the observed energies of $^{6}$He and is applied to the proton-rich nuclei in this analysis.

\subsection{Complex scaling method (CSM)}

We explain CSM, which describes resonances and nonresonant continuum states \cite{ho83,moiseyev98,aoyama06}.
Hereafter, we refer to the nonresonant continuum states as the continuum states simply. 
In CSM, we transform the relative coordinates of the $^4$He+$N_{\rm v} p$ system, as $\vc{r}_i \to \vc{r}_i\, e^{i\theta}$
for $i=1,\cdots,N_{\rm v}$, where $\theta$ is a scaling angle.
The Hamiltonian in Eq.~(\ref{eq:Ham}) is transformed into the complex-scaled Hamiltonian $H_\theta$, and the corresponding complex-scaled Schr\"odinger equation is given as
\begin{eqnarray}
	H_\theta\Psi^J_\theta
&=&     E\Psi^J_\theta .
	\label{eq:eigen}
\end{eqnarray}
The eigenstates $\Psi^J_\theta$ are obtained by solving the eigenvalue problem of $H_\theta$ in Eq.~(\ref{eq:eigen}).
In CSM, we obtain all the energy eigenvalues $E$ of bound and unbound states on a complex energy plane, governed by the ABC theorem \cite{ABC}.
In this theorem, it is proved that the boundary condition of resonances is transformed to one of the damping behavior at the asymptotic region.
This proof is mathematically a general one for the many-body system including a long range interaction.
The transformed boundary condition for the resonances makes it possible to use the same method to obtain the bound states and resonances.
This property of CSM is worthy to obtain the wave functions of the resonances directly as the Gamow states, in particular, for many-body case.
For a finite value of $\theta$, every Riemann branch cut starting from the different thresholds is commonly rotated down by $2\theta$.
Hence, the continuum states such as $^7$B+$p$ and $^6$Be+2$p$ channels in $^8$C are obtained on the branch cuts rotated by the $-2\theta$ 
from the corresponding thresholds \cite{myo077,myo097}.
In contrast, bound states and resonances are obtainable independently of $\theta$.
We can identify the resonance poles with complex eigenvalues: $E=E_r-i\Gamma/2$, where $E_r$ and $\Gamma$ are the resonance energies and the decay widths, respectively. 
In the wave function, the $\theta$ dependence is included in the expansion coefficients in Eq.~(\ref{WF0}) as $\{C_c^J(\theta)\}$. 
The coefficients $C_c^J(\theta)$ can be a complex number in general for a finite angle $\theta$.
The angle $\theta$ is determined to search for the stationary point of each resonance in a complex energy plane\cite{aoyama06,ho83,moiseyev98}.

The resonant state generally has a divergent behavior at asymptotic distances, and then its norm is defined by a singular integral using, 
for example, the convergent factor method\cite{aoyama06,romo68,homma97}.
In CSM, on the other hand, resonances are precisely described as eigenstates expanded in terms of the $L^2$ basis functions.
The amplitudes of the resonances are finite and normalized as $\sum_{c} \left(C_c^J(\theta)\right)^2=1$.
The Hermitian product is not applied due to the bi-orthogonal relation \cite{ho83,moiseyev98,berggren68}.
The matrix elements of resonances are calculated using the amplitudes obtained in CSM and are independent of the angle $\theta$ \cite{aoyama06}.

In this study, we discretize the continuum states in terms of the basis expansion, as shown in the figures of energy eigenvalue distributions in Refs. \cite{myo01,myo097,aoyama06}.
The reliability of the continuum discretization in CSM has already been shown using the continuum level density\cite{suzuki05} and the phase shift analysis\cite{kruppa07}.

\section{Results}\label{sec:result}

\subsection{Energy spectra of $^5$Li, $^6$Be, $^7$B and $^8$C}

\begin{figure}[t]
\centering
\includegraphics[width=8.5cm,clip]{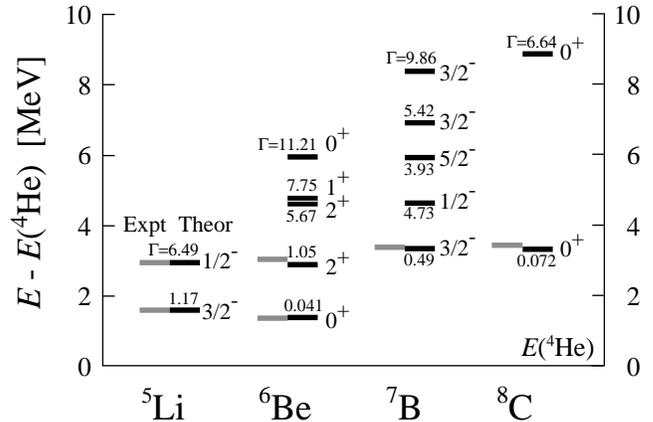}
\caption{Energy levels of $^5$Li, $^6$Be, $^7$B and $^8$C measured from the $^4$He energy. Units are in MeV.
Black and gray lines are theory and experiments, respectively. Small numbers are decay widths.}
\label{fig:C8}
\end{figure}

We show the systematic behavior of level structures of $^5$Li, $^6$Be, $^7$B and $^8$C in Fig. \ref{fig:C8}. There is no bound states in those nuclei.
It is found that the present calculations agree with the observed energy levels. We furthermore predict many resonances for $^6$Be, $^7$B and $^8$C.
In the previous analysis\cite{myo117}, we discussed the structures of $^6$Be and $^7$B in detail, such as the spatial properties of extra protons, the configurations and the mirror symmetry. 
It was found that the $^6$Be structures are similar to those of a mirror nucleus $^6$He.
For $^7$B, only the ground states of $^7$B and $^7$He breaks the mirror symmetry, while the excited states of two nuclei retain the symmetry\cite{myo117}.

\begin{table}[b]
\caption{Energy eigenvalues of the $^8$C ($0^+$) resonances measured from the $^4$He+$4p$ threshold.
The values with parentheses are the experimental ones\cite{charity11}. }
\label{ene8}
\centering
\begin{ruledtabular}
\begin{tabular}{c|ccc}
          & Energy~[MeV]      &  Width~[MeV]       \\ \hline
 $0^+_1$  & 3.32~[3.449(30)]  &  0.072~[0.130(50)] \\
 $0^+_2$  & 8.88              &  6.64              \\
\end{tabular}
\end{ruledtabular}
\end{table}

In this analysis, we discuss the structures of the $0^+$ states of $^8$C.
The energy eigenvalues are listed in Table \ref{ene8} measured from the $^4$He+4$p$ threshold.
We obtain two resonances of $^8$C($0^+$), both of which are five-body resonances as shown in Fig.~\ref{fig:C8}.
The energy of the $^8$C ground state is obtained as $E_r$=3.32 MeV and agrees with the recent experiment of $E_r=3.449(30)$ MeV\cite{charity11}.
The decay width is 0.072 MeV, which is small and good but slightly smaller than the experimental value of 0.130(50) MeV.
There is no experimental evidence for the excited states of $^8$C so far, and further experimental data are anticipated.

\begin{table}[t]
\caption{Dominant parts of the complex squared amplitudes $(C^J_c)^2$ of the ground states of $^8$C and $^8$He.}
\label{comp8_1}
\centering
\begin{ruledtabular}
\begin{tabular}{c|ccc}
Configuration              & $^8$C($0^+_1$)  &  $^8$He($0^+_1$) \\ \hline
 $(p_{3/2})^4$             & $0.878-i0.005$  &  0.860  \\
 $(p_{3/2})^2(p_{1/2})^2$  & $0.057+i0.001$  &  0.069  \\
 $(p_{3/2})^2(1s_{1/2})^2$ & $0.010+i0.003$  &  0.006  \\
 $(p_{3/2})^2(d_{3/2})^2$  & $0.007+i0.000$  &  0.008  \\
 $(p_{3/2})^2(d_{5/2})^2$  & $0.037+i0.000$  &  0.042  \\
 other 2$p$2$h$            & $0.008+i0.000$  &  0.011  \\
\end{tabular}
\end{ruledtabular}
\end{table}

We discuss the configuration properties of two resonances of $^8$C in detail.
For the ground state, in Table \ref{comp8_1}, we list the main configurations with their complex squared amplitudes $(C^J_c)^2$ in Eq. (\ref{WF0}).   
In general, the squared amplitude of a resonant state can be a complex number, while the total of the complex squared amplitudes is normalized as unity. 
The interpretation of the complex number in the physical quantity of resonances is still an open problem\cite{homma97}. 
In the results of $^8$C, the amplitudes of the dominant configurations are almost real values.
In that case, it is reasonable to discuss the physical meaning of the dominant real part of the amplitudes of the resonances in the same way that we discuss the bound state.
When we consider all the available configurations, the summations conserve unity due to the normalization of the states.

From Table \ref{comp8_1}, in the $^8$C ground state, the $(p_{3/2})^4$ configuration dominates the total wave function with a squared amplitude of 0.88 for real part.
The $2p2h$ excitations from the lowest $p_{3/2}$ orbit are mixed totally by about 0.12.
In the $2p2h$ components, the $p_{1/2}$ and $d_{5/2}$ orbits are rather contributing to the ground state.
These results mean that the $jj$ coupling scheme and the $p_{3/2}$ sub-closed nature are well established in the ground state of $^8$C.
To see the mirror symmetry, the results of the $^8$He ground state described in the $^4$He+$4n$ model\cite{myo10} are shown in Table \ref{comp8_1}.
In the $^8$He ground state, extra four neutrons dominantly occupy the $p_{3/2}$ orbit with a squared amplitude of 0.86
and the $2p2h$ components are mixed by about 0.14.
From those values of the squared amplitudes, it is concluded that the trend of the configuration mixing in the ground states of $^8$C and $^8$He is quite similar,
which indicates the good mirror symmetry between two states.
It is also noticed that among the $2p2h$ components, the only $(p_{3/2})^2(1s_{1/2})^2$ configuration of $^8$C increases slightly from that of $^8$He.
This is considered to be so-called the Thomas-Erhman shift caused by the Coulomb repulsion.

\begin{table}[t]  
\caption{Radial properties of the ground states of $^8$C and $^8$He in units of fm,
in comparison with the experiments of $^8$He; a\cite{tanihata92}, b\cite{alkazov97}, c\cite{kiselev05}, d\cite{mueller07}.}
\label{radius}
\centering
\begin{ruledtabular}
\begin{tabular}{c|cccc}
                        & $^8$C        & $^8$He    & $^8$He(exp.) \\ \hline
$R_{\rm m}$             & $2.81-i$0.08 &~~~2.52~~~ &  2.49(4)$^{\rm a}$,~2.53(8)$^{\rm b}$,~2.49(4)$^{\rm c}$   \\
$R_p$                   & $3.06-i$0.10 & 1.80      &  \\
$R_n$                   & $1.90-i$0.01 & 2.72      &  \\
$R_{\rm ch}$            & $3.18-i$0.09 & 1.92      &  1.929(26)$^{\rm d}$ \\
$r_{{\rm c}\mbox{-}4N}$ & $2.36-i$0.03 & 2.05      &   \\
\end{tabular}
\end{ruledtabular}
\end{table}

The radial properties of $^8$C are interesting to discuss the effect of the Coulomb repulsion in comparison with $^8$He having a neutron skin structure, 
although the radius of $^8$C can be complex numbers because of the resonance.
The results of $^8$C are shown in Table \ref{radius} for matter ($R_{\rm m}$), proton ($R_p$), neutron($R_n$) charge ($R_{\rm ch}$) parts,
and the relative distances between the $^4$He core and the center of mass of four valence nucleons ($r_{{\rm c}\mbox{-}4N}$).
It is found that the values in $^8$C are almost real, so that the real parts can be considered to represent the radial properties of $^8$C.
The matter radius of $^8$C is larger than that of $^8$He by about 11\% for a real part.
The relative distance between the $^4$He core and $4p$ in $^8$C is wider than the one between the $^4$He core and $4n$ in $^8$He by about 15\%. 
The enhancement of the radius of $^8$C from $^8$He comes from the Coulomb repulsion between five constituents of $^4$He+$p$+$p$+$p$+$p$ in $^8$C.
The Coulomb repulsion makes the energy of $^8$C shift up to become a resonance in comparison with $^8$He, 
and it also increases the relative distances between each constituent from the neutron skin state of $^8$He.

\begin{table}[t]
\caption{Dominant parts of the complex squared amplitudes $(C^J_c)^2$ of the $0^+_2$ states of $^8$C and $^8$He.}
\label{comp8_2}
\centering
\begin{ruledtabular}
\begin{tabular}{c|ccc}
Configuration              & $^8$C($0^+_2$)  &  $^8$He($0^+_2$) \\ \hline
 $(p_{3/2})^4$             & $0.044+i0.007$  &  $0.020-i0.009$  \\
 $(p_{3/2})^2(p_{1/2})^2$  & $0.934-i0.012$  &  $0.969-i0.011$  \\
 $(p_{3/2})^2(1s_{1/2})^2$ & $-0.001+i0.000$ & $-0.010-i0.001$  \\
 $(p_{3/2})^2(d_{3/2})^2$  & $0.020+i0.003$  &  $0.018+i0.022$  \\
 $(p_{3/2})^2(d_{5/2})^2$  & $0.002+i0.001$  &  $0.002+i0.000$  \\
\end{tabular}
\end{ruledtabular}
\end{table}

\begin{figure}[t]
\centering
\includegraphics[width=8.5cm,clip]{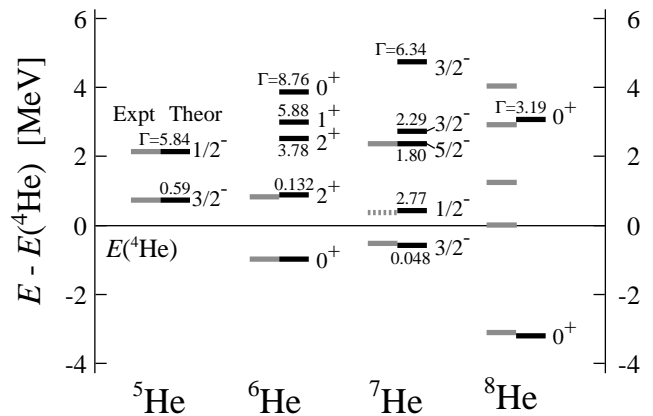}
\caption{Energy levels of He isotopes measured from the $^4$He energy. Units are in MeV.
Black and gray lines are theory and experiments, respectively. Small numbers are decay widths.
For $^7$He($1/2^-$), the reference experimental data is taken from Ref. \cite{skaza06} with gray dotted line.
For $^8$He, the experimental data are taken from Ref. \cite{golovkov09}, and only the $0^+$ states are shown in theory.}
\label{fig:He8}
\end{figure}

We discuss the excited $0^+_2$ state of $^8$C, which is located at the excitation energy of 5.6 MeV.
The dominant configurations of four valence protons are listed in Table \ref{comp8_2}.
In this state, the $(p_{3/2})^2(p_{1/2})^2$ configuration dominates the total wave function with a large squared amplitude of 0.93 for a real part,
while $(p_{3/2})^4$ is given as 0.04.
Hence, the $0^+_2$ state of $^8$C corresponds to the $2p2h$ excited state of the ground state
and can be described mostly in terms of the single configuration rather than the ground state.
This $2p2h$ configuration property is commonly seen in the $^8$He($0^+_2$) \cite{myo10} as shown in Table \ref{comp8_2}.
The coupling properties of four valence protons in $^8$C are discussed from the viewpoint of the proton pair numbers later.

\begin{figure}[t]
\centering
\includegraphics[width=8.5cm,clip]{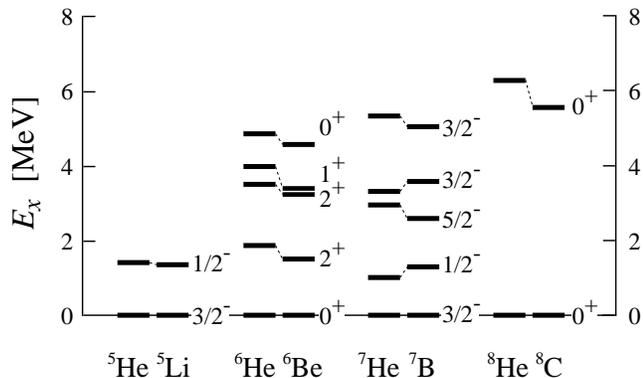}
\caption{Excitation energy spectra of mirror nuclei of $A=5,6,7$ and 8 in the units of MeV.}
\label{fig:excite}
\end{figure}

It is interesting to discuss the mirror symmetry between $^8$C and $^8$He consisting of $^4$He and four valence protons or neutrons.
To do this, we show the energy spectra of He isotopes with COSM in Fig.~\ref{fig:He8}, 
using the Hamiltonian in Eq.~(\ref{eq:Ham}) without the Coulomb term. 
For $^7$He(1/2$^-$), the experimental energy is not fixed, so we include the recent data \cite{skaza06} with dotted line as a reference in the figure.
From the figure, it is found that the COSM results agree with the observed energy levels well for He isotopes.
From Figs.~\ref{fig:C8} and \ref{fig:He8}, it is found that the orders of energy levels are the same between proton-rich and neutron-rich sides.
In the proton-rich side, the whole spectra are shifted up due to the Coulomb repulsion, in comparison with those of the neutron-rich side.
The displacement energies are 2.4 MeV for $^6$Be from $^6$He, 3.9 MeV for $^7$B from $^7$He, and 6.5 MeV for $^8$C from $^8$He, respectively.
In Fig. \ref{fig:excite}, we compare the excitation energy spectra of proton-rich and neutron-rich sides.
It is found that the good symmetry is confirmed between the corresponding nuclei.
The differences of excitation energies for individual levels are less than 1 MeV.

\subsection{Occupation and pair numbers in $^8$C}\label{sec:pair}

\begin{table}[t]
\caption{Complex occupation numbers of valence protons in $^8$C.}
\label{tab:occupy}
\centering
\begin{ruledtabular}
\small
\begin{tabular}{p{1.0cm}|p{2.2cm} p{2.2cm}}
Orbit      &~~~~$0^+_1$      &~~~~$0^+_2$        \\ 
\hline
 $p_{1/2}$ &~~$0.12+i 0.00$  &~~ $1.87 -i 0.02$  \\
 $p_{3/2}$ &~~$3.75-i 0.01$  &~~ $2.09 +i 0.01$  \\
 $s_{1/2}$ &~~$0.03+i 0.00$  &~~ $0.00 +i 0.00$  \\
 $d_{3/2}$ &~~$0.02+i 0.00$  &~~ $0.04 +i 0.01$  \\
 $d_{5/2}$ &~~$0.08+i 0.00$  &~~ $0.01 -i 0.00$  \\
\end{tabular}
\end{ruledtabular}
\end{table}

\begin{table}[t]
\caption{Complex pair numbers $P(J^\pi,S)$ of valence protons in $^8$C up to $J^\pi=3^-$.}
\label{tab:pair}
\centering
\begin{ruledtabular}
\small
\begin{tabular}{p{0.3cm}p{0.3cm}|p{2.2cm} p{2.2cm}}
$J^\pi$ &$S$ &~~~~$0^+_1$        &~~~~$0^+_2$       \\ 
\hline
 $0^+$  & 0  &~~$0.69 + i 0.00$  &~~$0.98 + i 0.00$  \\
 $0^+$  & 1  &~~$0.37 + i 0.00$  &~~$0.96 + i 0.00$  \\
\hline
 $0^-$  & 0  &~~$0.00 + i 0.00$  &~~$0.00 + i 0.00$  \\
 $0^-$  & 1  &~~$0.00 + i 0.00$  &~~$0.01 + i 0.00$  \\
\hline
 $1^+$  & 0  &~~$0.01 + i 0.00$  &~~$0.15 + i 0.00$  \\
 $1^+$  & 1  &~~$0.08 + i 0.00$  &~~$1.24 - i 0.02$  \\
\hline
 $1^-$  & 0  &~~$0.02 + i 0.00$  &~~$0.00 + i 0.00$  \\
 $1^-$  & 1  &~~$0.02 + i 0.00$  &~~$0.02 + i 0.00$  \\
\hline
 $2^+$  & 0  &~~$1.54 + i 0.00$  &~~$0.86 + i 0.00$  \\
 $2^+$  & 1  &~~$3.06 - i 0.01$  &~~$1.72 + i 0.00$  \\
\hline
 $2^-$  & 0  &~~$0.02 + i 0.00$  &~~$0.00 + i 0.00$  \\
 $2^-$  & 1  &~~$0.07 + i 0.01$  &~~$0.01 + i 0.00$  \\
\hline
 $3^+$  & 0  &~~$0.00 + i 0.00$  &~~$0.00 + i 0.00$  \\
 $3^+$  & 1  &~~$0.00 + i 0.00$  &~~$0.00 + i 0.00$  \\
\hline
 $3^-$  & 0  &~~$0.04 + i 0.00$  &~~$0.02 + i 0.00$  \\
 $3^-$  & 1  &~~$0.09 + i 0.00$  &~~$0.02 + i 0.00$  \\
\end{tabular}
\end{ruledtabular}
\end{table}

We discuss the structures of the two $0^+$ states of $^8$C from the view point of the configurations of valence protons.
We list the complex occupation numbers of four protons in each orbit for $^8$C in Table \ref{tab:occupy}.
Since the states are Gamow states, their physical quantities become complex values with a relatively small imaginary part,
while their summation conserves the valence proton number being four as a real value in each state.
In the $^8$C ground state, the $p_{3/2}$ orbit is dominant and its real part of the complex occupation number of 3.75 is close to four,
which is consistent with the dominant configuration of $(p_{3/2})^4$ as shown in Table \ref{comp8_1}.
In the $0^+_2$ state, the $p_{3/2}$ and $p_{1/2}$ orbits share two protons individually
because the $(p_{3/2})^2(p_{1/2})^2$ configuration dominates this state as shown in Table \ref{comp8_2}.
In two $0^+$ states, the interaction acting between valence protons emerges the small mixing of the other orbits such as the $sd$ shell components.

We calculate the complex pair number $P({J^\pi},S)$ of four valence protons in $^8$C, which is defined by the matrix element
of the operator as
\begin{eqnarray}
P({J^\pi},S) &=& \langle\sum_{\alpha \le \beta} A^\dagger_{J^\pi,S}(\alpha\beta)A_{J^\pi,S}(\alpha\beta)\rangle.
\end{eqnarray}
Here the quantum numbers $\alpha$ and $\beta$ are for the single particle state, and 
$A^\dagger_{J^\pi,S}$ ($A_{J^\pi,S}$) is the creation (annihilation) operator of a proton pair with the coupled angular momentum and parity $J^\pi$ and the coupled intrinsic spin $S$.
The complex pair numbers are useful to understand the structures of four protons from the viewpoint of pair coupling.
The summation of the complex pair numbers over all $J^\pi$ and $S$ is equal to six, which is a real value, from the total pair number consisting of four protons as
\begin{eqnarray}
\sum_{J^\pi,S} P({J^\pi},S) &=& 6.
\end{eqnarray}

\begin{figure}[t]
\centering
\includegraphics[width=7.5cm,clip]{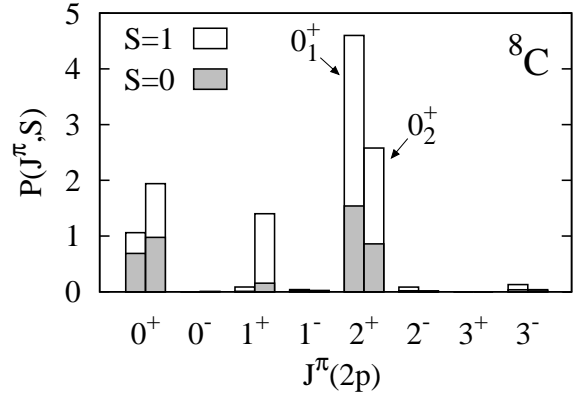}
\caption{Real parts of the complex pair numbers of valence protons $P({J^\pi},S)$ in the $^8$C($0^+_1$, $0^+_2$) states decomposed into the $S=0$ (shaded) and $S=1$ components (blank).}
\label{fig:pair_8C}
\end{figure}

In Table \ref{tab:pair}, we list the results of the complex pair numbers up to the $3^-$ component for two $0^+$ states of $^8$C.
It is found that the values are almost real and their imaginary part is very small and hence we focus on the discussion of the results of real parts. 
In Fig.~\ref{fig:pair_8C}, we show the real part of the complex pair numbers and compare them between two $0^+$ state of $^8$C.
In the ground state, it is found that the $2^+$ proton pair is dominant with the real part of the number as about 4.6 taking the summation of $S=0$ and $1$, and the $0^+$ proton pair number is about 1.1 for a real part.
These results are consistent with the main configuration of $(p_{3/2})^4$ from the CFP decomposition with the numbers of 1 and 5 for the $0^+$ and $2^+$ pairs, respectively.
The decompositions into the $S=0$ and $S=1$ components can also be naively understood from the $(p_{3/2})^2_{J=0,2}$ configuration using the $LS$ coupling transformation.
These results show that the shell structure of the $p_{3/2}$ protons is well established in the $^8$C ground state from the pair numbers.
When one of the proton pair with $J^\pi=0^+$ or $2^+$ state is coupled with $^4$He, this system can be the main components of $^6$Be($0^+_1$) and $^6$Be($2^+_1$), respectively \cite{myo117}.
Recent experiment\cite{charity10,charity11} shows that the decay of the $^8$C ground state can go through the $^6$Be($0^+_1$)+$2p$ channel while the final states are the five-body $^4$He+$4p$ system.
When they reconstruct the $^6$Be($0^+_1$) component among the five-body final states, the probability of the $^6$Be($0^+_1$)+2$p$ decay channel is estimated as 0.92(5),
which is consistent to the present value of the $0^+$ pair number summed by the spin $S$ shown in Table \ref{tab:pair}.

In the $0^+_2$ state of $^8$C, this state has about 2.0 of the $0^+$ proton pair number for a real part in addition to the large $2^+$ pair number as about 2.6.
This is consistent with the $(p_{3/2})^2(p_{1/2})^2$ configuration,
which can be decomposed into the pairs of $0^+$, $1^+$ and $2^+$ with the numbers of $2$, $1.5$ and $2.5$, respectively. 
Similarly to the ground state case, the decompositions into the $S=0,1$ components in the one proton-pair are naively understood from the $(p_{3/2})^2$ and $(p_{1/2})^2$ configurations.

\begin{figure}[t]
\centering
\includegraphics[width=7.5cm,clip]{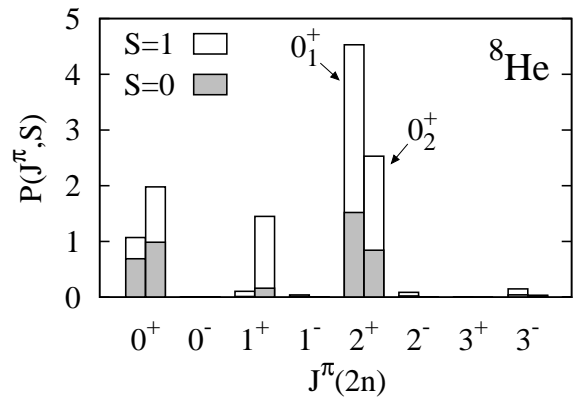}
\caption{Real parts of the complex pair numbers of valence neutrons $P({J^\pi},S)$ in the $^8$He($0^+_1$, $0^+_2$) states decomposed into the $S=0$ (shaded) and $S=1$ components (blank).}
\label{fig:pair_8He}
\end{figure}

To discuss the mirror symmetry of $^8$C, we show the neutron pair numbers of four valence neutrons in $^8$He in Fig.~\ref{fig:pair_8He}, those of which dominantly have the real parts.
From the results, it is found the distributions are quite similar between $^8$C and $^8$He for the ground and excited $0^+$ states.
Similar to the case of $^8$C, the importance of the $2^+$ neutron pair in $^8$He is confirmed.
This result was suggested in the experiment \cite{korsheninnikov03}, and is also obtained in the $^6$He+$n$+$n$ three-body analysis \cite{adahchour06}.
On the other hand, the $0^+_2$ state has almost two of the $0^+$ neutron pair number in addition to the large $2^+$ pair number.
This is consistent with the $(p_{3/2})^2(p_{1/2})^2$ configuration, as shown in Table \ref{comp8_2}.
In summary, the structures of $^8$C are similar to those of $^8$He for the properties of the pair numbers of valence nucleons above $^4$He.  
This result indicates that the mirror symmetry is well retained in two nuclei for the $0^+$ states.

\section{Summary}\label{sec:summary}

We have investigated the resonance structures of $^8$C with the $^4$He+$p+p+p+p$ five-body cluster model.
The boundary condition for many-body resonances is accurately treated using the complex scaling method. 
The decay thresholds concerned with subsystems are described consistently.
We have found two $0^+$ resonances of $^8$C, which are five-body resonances and are dominantly described by the $p$-shell configurations. 
For the ground state, the energy and the decay width agree with the recent new experiments.
We also predict the excited $0^+_2$ resonance of $^8$C, which we hope to see confirmed experimentally.
It is found that the present cluster model describes well the systematic energy spectra of proton-rich nuclei from $^5$Li to $^8$C,
in addition to the mirror nuclei of the neutron-rich He isotopes from $^5$He to $^8$He.

For $^8$C, we furthermore investigate the structures of four valence protons around the $^4$He core and compare them with those of neutrons in $^8$He, a mirror nucleus.
The ground state of $^8$C is dominated by the $(p_{3/2})^4$ configuration of four protons with the squared amplitude of about 0.88 and the $0^+_2$ state is $(p_{3/2})^2(p_{1/2})^2$ with about 0.93, corresponding to the $2p2h$ configuration from the ground state. 
We also decompose the four protons into two proton-pairs and discuss the coupling behavior of the two proton-pairs.
It is found that the $2^+$ proton pair contributes largely in the $^8$C ground state which is understood from the $(p_{3/2})^4$ configuration.
On the other hand, the $^8$C($0^+_2$) state has about two of the $0^+$ proton pairs which mainly comes from the $(p_{3/2})^2(p_{1/2})^2$ configuration.
The structure of $^8$C is compared with $^8$He and it is found that both the ground and excited $0^+$ states of two nuclei retain the mirror symmetry well.

\section*{Acknowledgments}
This work was supported by a Grant-in-Aid for Young Scientists from the Japan Society for the Promotion of Science (No. 21740194).
Numerical calculations were performed on a supercomputer (NEC SX9 and SX8R) at RCNP, Osaka University.

\section*{References}
\def\JL#1#2#3#4{ {{\rm #1}} \textbf{#2}, #4 (#3)}  
\nc{\PR}[3]     {\JL{Phys. Rev.}{#1}{#2}{#3}}
\nc{\PRC}[3]    {\JL{Phys. Rev.~C}{#1}{#2}{#3}}
\nc{\PRA}[3]    {\JL{Phys. Rev.~A}{#1}{#2}{#3}}
\nc{\PRL}[3]    {\JL{Phys. Rev. Lett.}{#1}{#2}{#3}}
\nc{\NP}[3]     {\JL{Nucl. Phys.}{#1}{#2}{#3}}
\nc{\NPA}[3]    {\JL{Nucl. Phys.}{A#1}{#2}{#3}}
\nc{\PL}[3]     {\JL{Phys. Lett.}{#1}{#2}{#3}}
\nc{\PLB}[3]    {\JL{Phys. Lett.~B}{#1}{#2}{#3}}
\nc{\PTP}[3]    {\JL{Prog. Theor. Phys.}{#1}{#2}{#3}}
\nc{\PTPS}[3]   {\JL{Prog. Theor. Phys. Suppl.}{#1}{#2}{#3}}
\nc{\PRep}[3]   {\JL{Phys. Rep.}{#1}{#2}{#3}}
\nc{\JP}[3]     {\JL{J. of Phys.}{#1}{#2}{#3}}
\nc{\andvol}[3] {{\it ibid.}\JL{}{#1}{#2}{#3}}

\end{document}